\documentclass[twocolumn]{aastex61}
\usepackage{mathrsfs,amssymb,amsmath}
\hypersetup{breaklinks=true}
\usepackage{breakcites}

\newcommand\aastex{AAS\TeX}

\submitjournal{ApJ}

\shorttitle{\aastex\ Electrostatic barrier}
\shortauthors{Akimkin et al.}

\begin{document}

\title{Inhibited coagulation of micron-size dust due to the electrostatic barrier}

\correspondingauthor{Vitaly Akimkin} \email{akimkin@inasan.ru}

\author[0000-0002-4324-3809]{V.V. Akimkin}
\affil{Institute of Astronomy, Russian Academy of Sciences, Pyatnitskaya str. 48, Moscow, 119017, Russia}

\author[0000-0002-1590-1018]{A.V. Ivlev}
\affiliation{ Max-Planck-Institut f\"ur Extraterrestrische Physik, Giessenbachstr. 1, Garching, D-85748, Germany}

\author[0000-0003-1481-7911]{P. Caselli}
\affiliation{ Max-Planck-Institut f\"ur Extraterrestrische Physik, Giessenbachstr. 1, Garching, D-85748, Germany}

\begin{abstract}
The collisional evolution of solid material in protoplanetary disks is a crucial step in the formation of planetesimals,
comets, and planets. Although dense protoplanetary environments favor fast dust coagulation, there are several factors that
limit the straightforward pathway from interstellar micron-size grains to pebble-size aggregates. Apart from the grain
bouncing, fragmentation, and fast drift to the central star, a notable limiting factor is the electrostatic repulsion of
like-charged grains. In this study we aim at theoretical modeling of the dust coagulation coupled with the dust charging and
disk ionization calculations. We show that the electrostatic barrier is a strong restraining factor to the coagulation of
micrometer-size dust in dead zones of the disk (where the turbulence is suppressed). While the sustained turbulence helps to
overcome the electrostatic barrier, low fractal dimensions of dust aggregates can potentially block their further
coagulation even in this case. Coulomb repulsion may keep a significant fraction of small dust in the disk atmosphere and
outer regions.

\end{abstract}

\keywords{accretion, accretion disks  --- dust, extinction --- meteorites, meteors, meteoroids --- plasmas --- protoplanetary disks
--- turbulence }

\section{Introduction} \label{sec:intro}

Grain charge is involved in various aspects of cosmic dust physics. It affects dust drift through ionized
gas~\citep{1979ApJ...231...77D, 1979A&A....77..165G, 2017MNRAS.469..630A,2018MNRAS.473.1576K}, dust interaction with magnetic
fields~\citep{1977A&A....55..253S, 1980ApJ...241.1021D, 2017MNRAS.469.3532L, 2018MNRAS.479.4681H}, gas
ionization-recombination balance and chemistry \citep{1980PASJ...32..405U, 2000ApJ...543..486S, 2006A&A...445..205I,
2013ApJ...765..114D, 2016ApJ...833...92I, 2018Ap&SS.363..151N, 2019A&A...632A..44T}. Particle charge state influences the
radiative properties of dust, specifically the interaction of electromagnetic waves with optically small
particles~\citep{1977CaJPh..55.1930B, 2012OptL...37..265K, 2012JQSRT.113.2561K} and infrared bands of polycyclic aromatic
hydrocarbons~\citep{1993ApJ...408..530D, doi:10.1021/jp952074g, 2018ApJ...858...67B}. Coulomb potential also changes the
collisional cross section with gas and solids, thus affecting dust evolution due to ion
accretion~\citep{1967ApJ...147..965M, 1999ApJ...517..292W, 2018ApJ...857...94Z} and coagulation~\citep{1978Ap&SS..57..381S,
1979Ap&SS..61...65S, 2009ApJ...698.1122O, 2013ApJ...776..103M, 2015ARep...59..747A}. Betatron acceleration of charged grains
leads to their sputtering and shuttering in high velocity shock waves~\citep{1987ApJ...318..674M, 1994ApJ...431..321T,
1996ApJ...469..740J}. Moreover, extreme values of charge may lead to the dust destruction via the ion field emission or
Coulomb explosions~\citep{1979ApJ...231...77D, 2000ApJ...537..796W, 2011AIPC.1397..249K}.

The study of cosmic dust charging has a long history~\citep{1937AN....263..425J, 1938ZA.....15..239C, 1941ApJ....93..369S,
1972ApJ...176..103W, 1973ApJ...181..101F} and resulted in general understanding of main charging mechanisms acting in the
interstellar medium~(see reviews by \citet{1989RvGeo..27..271G, 1994ARA&A..32..419M, 2005PhR...421....1F} and short summary by
\citet{2004ASPC..309..453W}). These mechanisms include photoelectric effect~\citep{1994ApJ...427..822B,
2001ApJS..134..263W}, plasma charging~\citep{1987ApJ...320..803D}, secondary electron emission~\citep{1979ApJ...231...77D,
2007A&AT...26..227S}, triboelectric charging\footnote{Both plasma charging and triboelectric charging are sometimes referred
to as collisional charging.} \citep{2000ApJ...533..472P, 2000Icar..143...87D, 2005GeoRL..3211202M,
2018PhRvE..97b2904S, 0004-637X-867-2-123} and thermionic emission~\citep{1975A&A....41..437L}. In varying external conditions
the non-equilibrium grain charging should be considered and may affect dust
evolution \citep{1938ZA.....15..239C, 1990ApJ...361..155H, 2011ApJ...740...77P}. However, even in steady-state conditions the
grain charge may fluctuate around an average value and an ensemble of equal size grains has non-zero charge dispersion.

The key charging mechanisms in protoplanetary disks are photoelectric, plasma, and triboelectric charging, any of which can
dominate depending on disk region or grain size. The charge of small dust is positive in disk atmosphere, due
to the photoelectric emission caused by the combined effect of the interstellar UV field and the UV radiation generated
locally by the penetrating cosmic rays (CRs; \citet{Ivlev2015}). In disk interiors, the plasma charging dominates and
grains are charged predominantly negatively. Grains may exchange their charges in mutual collisions; however, high dust
concentration and low ionization degree are needed for triboelectric charging to be dominating over the plasma and
photoelectric charging. Unlike the triboelectric charging, which is considered as a minor factor in the growth of dust in
protoplanetary disks~\citep{2010RAA....10.1199B}, the plasma and photoelectric charging plays a major role in the
collisional evolution of $\sim1\,\mu$m grains~\citep{2011ApJ...731...95O, 2011ApJ...731...96O, 2015ARep...59..747A}.

As the grain charge typically scales linearly with the grain size, the electrostatic repulsion of like-charged grains become
stronger as dust grows. At the same time, kinetic energy in Brownian motions does not depend on grain mass, so the purely
thermal dust coagulation inevitably stops at some point due increasing Coulomb repulsion. Non-thermal motions, which can be
induced by turbulence or differential dust drift, may provide the necessary kinetic energy for dust grains to overcome the
electrostatic barrier. Subsonic turbulence present in dense cores~\citep{1998ApJ...504..207B,
2002ApJ...572..238C, 2011ApJ...739L...2P, 2015MNRAS.446.3731K} leads to a weak dependence of dust coagulation on grain
charging~\citep{1993ApJ...407..806C}, as the resulting kinetic energy of colliding grains is typically larger than their
repulsion energy. However, the notably weaker turbulence in protoplanetary disks may be insufficient to overcome the
electrostatic barrier.

\citet{2009ApJ...698.1122O} showed that the the electrostatic repulsion becomes important if the turbulence parameter
$\alpha= (v_{\rm turb}/c_{\rm s})^2$ is smaller than $\sim10^{-2}$, where $v_{\rm turb}$ is the mean turbulent velocity of
the gas and $c_{\rm s}$ is the sound speed. Recent observational efforts to constrain turbulence via CO line profiles in
protoplanetary disks put upper limits of $v_{\rm turb}<0.05c_{\rm s}$ and $<0.08c_{\rm s}$ for HD~163296 and TW~Hya,
respectively~\citep{2017ApJ...843..150F, 2018ApJ...856..117F}, which translates to $\alpha\lesssim0.003-0.006$. The
effective $\alpha$-parameter in the dead zones with suppressed magnetorotational instability (MRI) may be even smaller,
which poses an important challenge for current understanding of the dust evolution in protoplanetary disks.

Grain charging is still frequently neglected in present theoretical models of dust evolution, even so it may be invoked for
the explanation of some observational properties of protoplanetary disks. First, it was shown that unconstrained dust
coagulation leads to the depletion of small grains at timescales much shorter than the lifetimes of protoplanetary
disks~\citep{2005A&A...434..971D}. The fast coagulation of small dust would result in a significant drop in near- and mid-IR
fluxes, which contradicts to observations~\citep{2001AJ....121.1512H, 2007ApJ...667..308C}, so there should be a mechanism of
either the replenishment of small dust population (e.g., via fragmentation in high-speed collisions) or the slowdown of its
coagulation. The electrostatic barrier is a viable mechanism for such a slowdown as it can completely block the coagulation
of micron-size dust, especially in the disk atmosphere and outer regions. Second, as the electrostatic barrier is most
important for $\sim0.1-10\,\mu$m grains, it may divide dust population into small and large sub-populations.

In this paper we present numerical simulations of charged dust coagulation for typical protoplanetary disk conditions. These
simulations are done in 2D in radial and vertical disk extent, accounting for the co-evolving dead zone with suppressed
turbulence and tackles non-compact (fractal) grains. The present model does not consider global dust dynamics as well as
fragmentation, as these factors are important for macroscopic dust and are of lesser importance for micron-size grains, for
which the electrostatic barrier is crucial.

\section{Model description} \label{sec:model}
To study the grain charge impact on the dust evolution we solve both the coagulation equation and grain charging balance
equations. The corresponding model was initially presented in \citet{2015ARep...59..747A} and \citet{2017ASPC..510...63A},
here we recap its basic features and describe new improvements. The goal of these numerical simulations is to understand at
which conditions the electrostatic barrier operates and how it can be overcome.

The electrostatic barrier against dust growth is important for $0.1-10 \,\mu$m grains~\citep{2009ApJ...698.1122O}. Grains
of this size range do not experience substantial drift relative to the gas. Hence, in the basic approach, we treat the
problem locally, i.e. neglecting the possible influx and outflux of large drifting grains in/from the grid cell. In
Section~\ref{sec:results} we study how the presence of non-locally grown dust affects the problem.

The background physical conditions during the whole simulation run of 0.9\,Myr are assumed to be stationary. This
includes stellar parameters, disk density, and temperature distributions, but not the ionization degree as it could have an
important feedback loop with the charged dust evolution. The estimates of the protoplanetary disks lifetimes
\citep{2001ApJ...553L.153H, 2006ApJ...638..897S, 2014ApJ...793L..34P} are a factor of two to ten longer than our simulation
run of $0.9$~Myr, so we keep the global disk structure fixed to preserve the clarity in the study. The assumed central star
mass, radius, and effective temperature are $M_{\star}=0.7~M_{\odot}, R_{\star}=2.64~R_{\odot}, T_{\rm eff}=4000$\,K. 
The disk is assumed to be azimuthally symmetric with the radial profile of the gas surface density set by the power law tappered at the inner and outer characteristic radii $R_{\rm c}^{\rm inn}$ and $R_{\rm c}^{\rm out}$:
\begin{equation}
\Sigma(R)=\Sigma_0\left(\frac{R}{{\rm 1\,au}}\right)^{-\gamma} \exp\left[ -\left(\frac{R}{R_{\rm c}^{\rm out}}\right)^{2-\gamma} -\left(\frac{R}{R_{\rm c}^{\rm inn}}\right)^{\delta} \right].
\end{equation}
 While the values of $\gamma$ and $R_{\rm c}^{\rm out}$ can be estimated
from observations \citep{2011ARA&A..49...67W}, $\delta$ and $R_{\rm c}^{\rm inn}$ are loosely constrained. We adopt
$R_{\rm c}^{\rm inn}=0.5$\,au, $R_{\rm c}^{\rm out}=200$\,au, $\gamma=1$, and $\delta=-3$ with the normalization $\Sigma_0=300$\,g\,cm$^{-2}$, which results in 
the total gas mass $\approx0.02\,M_{\odot}$ within inner $10^3$\,au and $\Sigma({\rm 1\,au})=263$\,g\,cm$^{-2}$.
 The gas mass density $\rho_{\rm g}$ is calculated from the condition of the vertical hydrostatic equilibrium with (vertically) isothermal gas and
dust. For the temperatures we have $T_{\rm d}(R)=T_{\rm g}(R)=\varphi^{1/4} T_{\rm eff}(R/R_{\rm \star})^{-1/2}$, with $\varphi=0.05$ being the grazing
angle. This yields the radial scaling of midplane density $\rho_{\rm g}(R,0)\propto R^{-9/4}$. The dust-to-gas mass ratio is 0.01 for all disk locations and does not change with dust evolution, the grain solid
density $\rho_{\rm s}=3$ g cm$^{-3}$ and the initial size distribution follows the power law with boundaries $0.005\,\mu$m
and $0.25\,\mu$m and power-law slope $-3.5$.

The dust coagulation is modeled by the numerical solution of the Smoluchowski equation for charged grains. The coagulation
kernel for two colliding grains with radii $a_1, a_2$ and charges $Q_1(a_1), Q_2(a_2)$ can be written as
\begin{equation}\label{eq:coagulation_kernel}
  K_{12}(a_1,a_2)=\pi(a_1+a_2)^2u_{12}\mathcal{C}_{12},
\end{equation}
where
\begin{equation}\label{CF}
\mathcal{C}_{12}(a_1,a_2)=1-\frac{2Q_1Q_2}{(a_1+a_2)m_{12}u_{12}^2}
\end{equation}
is the Coulomb factor, $m_{12}$ is the reduced mass of two grains, and $u_{12}$ is their relative velocity. The additional
condition on the kernel is $\mathcal{C}\ge0$.
The dominant grain charging mechanisms for the selected physical conditions are plasma and photoelectric charging
\citep{1987ApJ...320..803D, 2001ApJS..134..263W}.
The former leads to the predominantly negative grain charge in the dark midplane, the latter leads to the positive dust
grains in the illuminated disk atmosphere. The competition of these two charging mechanisms produces a zero-charge surface
somewhere in the disk upper layers. It is important to consider the dispersion of grain charge, so $K_{12}$ should be
integrated over the possible charge states of a grain of a given size. We refer the reader to \citet{2015ARep...59..747A}
for more information on our approach to the Smoluchowski equation solution for charged dust. Below we describe the improvements
we made in the model in comparison with that work.

In dense dark conditions the number density of dust grains becomes sufficiently high to affect the overall charge balance,
leading to significant electron depletion onto grains~\citep{1980PASJ...32..405U, 2009ApJ...698.1122O}, so that electron and
ion number densities are no longer equal and $n_{\rm e}\ll n_{\rm i}$. This effect was not considered
in~\citet{2015ARep...59..747A}, so we added it following \citet{2016ApJ...833...92I}, where we studied the transition
between the electron--ion, dust--ion, and dust--dust plasma regimes. To find the local charge structure (distribution of dust
charges and abundance of free electrons and ions) one should solve the equations on overall charge neutrality and
ionization-recombination balance~\citep[see Equations~(11)--(12) in][]{2016ApJ...833...92I}. Input parameters for these
calculations were the dominant ion mass and the distribution of the total ionization rate $\zeta(R,z)$. The grain charge 
acquired due to collisions with plasma particles weakly depends on the ion mass. Heavy ions with $m_{\rm i}\gtrsim20\,m_{\rm H}$ 
are typically expected in the warm midplane regions, while lighter H$^+_3$
ions dominate in the colder outer parts~\citep{2004A&A...417...93S}. We adopt $m_{\rm i}=29\,m_{\rm H}$, which corresponds to
either HCO$^+$ or N$_2$H$^+$ ions, as the representative value for this paper.

CRs, X-rays, and radioactive elements are considered as main ionization sources. For CR ionization we use the approach
presented in~\citet[][]{2018A&A...614A.111P} (see their Equation (46) and appendix~F for model $\mathscr{H}$). As described in
\citet{2018A&A...614A.111P}, for the effective gas surface densities $\Sigma_{\rm eff}$ below the transition surface
density, $\Sigma_{\rm tr}\approx 130$~g~cm$^{-2}$, the ionization occurs mainly due to CR protons. Effective surface
density, which accounts for non-vertical magnetic field morphology, is taken to be 3.3 times the actual gas surface
density~\citep[see section 7.1 in][]{2018A&A...614A.111P}.\footnote{In many studies, the fact that CRs propagate along the
magnetic field is completely ignored. This generally leads to underestimated column density traversed by CRs and, hence, to
overestimated ionization.}  For regions with high surface densities, $\Sigma_{\rm eff}\gtrsim\Sigma_{\rm tr}$, the ionization
is dominated by secondary gamma-rays (produced by CRs) and, thus, the magnetic fields do not affect the results in this
case. In this case, the CR ionization rate should be calculated by the averaging over all direction to the transition surface; in
this work we do averaging only in the $z$-direction for simplicity.

For the X-ray ionization rate we adopt the approach from \citet{2009ApJ...701..737B} (see their Equation (21)) assuming stellar
X-ray luminosity $10^{30}$~erg~s$^{-1}$ and $T_X=3$~keV. The lowest total ionization rate is limited by the contribution of
radioactive decay, which we set to $1.4\times10^{-22}$~s$^{-1}$ \citep{2009ApJ...690...69U}.

The key part of any dust evolution model is the source of collisional velocities. The relative velocities due to the
Brownian motion $u_{\rm Br}$ of dust grains with the same size $a$ decrease with $a$. However the grain charge $Q$ typically
scales linearly with $a$, so the Coulomb factor in~Equation~\eqref{eq:coagulation_kernel} inevitably becomes zero for
sufficiently large $a$. Thus, the purely Brownian motion, although being the the main driver of early dust coagulation, can
not be responsible for the overcoming of the electrostatic barrier. However, it defines the critical grain size $\sim 1 \mu$m,
where dust coagulation starts to be strongly affected by the charge, which we studied in \citet{2015ARep...59..747A} and
\citet{2016ApJ...833...92I}. In this paper we also included the turbulence-induced collisional velocities using closed form
expressions from \citet{2007A&A...466..413O} using a procedure implemented by~\citet{2010A&A...513A..79B}. The stopping time
of a compact grain $t_{\rm s}=\rho_{\rm s} a/(v_{\rm th}\rho_{\rm g})$ \citep{Armitage2018} can be rewritten as
$t_{\rm s}=3 m(a)/(4v_{\rm th}\rho_{\rm g}\pi a^2)$ in the general case of fractal dust. Here, $m(a)$ is the
mass of the aggregate of size $a$ (consisting of monomers with material density $\rho_{\rm s}$) and $v_{\rm th}$ is the
thermal speed of molecules. The total collision velocity is $u_{12}=\sqrt{u_{\rm Br}^2+u_{\rm turb}^2}$. In our
simulations we account for the presence of a dead zone with suppressed turbulence and define it as a region with
sufficiently low ionization degree $x_{\rm e}<x_{\rm cr}=10^{-13}$~\citep{2014Ap&SS.352..103D}. We adopt the turbulence
parameter $\alpha_{\rm active}=10^{-3}$ for the MRI active region and $\alpha_{\rm dead}=10^{-6}$ for the dead zone. The
value of $\alpha_{\rm active}$ is taken according to the available observational constraints~\citep{2017ApJ...843..150F,
2018ApJ...856..117F}. As the gas ionization degree in dense regions depends on the current dust size distribution, we
recalculate the position of the dead zone consistently with the dust evolution.

\citet{2009ApJ...698.1122O} pointed out the importance of dust fluffiness in the evolution of charged dust. In our modeling
we consider several choices for the fractal dimension $D$ of dust aggregates. Fractal dimension $D$ is defined via a scaling
relation between aggregate mass $m$ and its characteristic size $a$,
\begin{equation} \label{eq:fractal}
 m(a)=m_0\left(\frac{a}{a_0}\right)^D,
\end{equation}
where $m_0=(4\pi/3)\rho_{\rm s}a_0^3$ is the mass of a compact monomer having radius $a_0$. For any choice of $D$ we assume
that initially all particles are compact, while all grown particles are made of monomers with the size equal to the largest
one in the initial distribution, i.e. $a_0=0.25\,\mu$m. To compare different cases of the fractal dimension $D$, we introduce
an equivalent radius $a_{\rm c}$ equal to the radius of a compact grain of the same mass
\begin{equation}
 a_{\rm c}^3=3m/(4\pi\rho_{\rm s})=a_0^{3-D}a^D.
\end{equation}

The charging of fractal dust aggregates is a complicated and poorly studied topic, as dipole interactions and stochastic
nature of interactions of free charge with asymmetrically charged aggregate are important factors. We neglect these
effects in the current study and assume that the cross section of a grain with mass $m$ is equal to $\pi a^2$, where $a$ is
defined from Equation~\eqref{eq:fractal}. This expression for cross section is used in both coagulation and charging
equations. The consideration of fractal dust is our third and last major improvement to the model
by~\citet{2015ARep...59..747A} (with electron depletion and turbulence-induced velocities being the other two ones).

\section{Results} \label{sec:results}
\subsection{Coagulation of charged grains}
The numerical simulations of charged dust coagulation provides us the evolution of the dust size distribution at a given
disk location. To show the broad picture of the grain charge impact on the dust growth, we compute the average grain radius in
different locations of 2D vertical cut of the disk and use it as a measure for the severity of the electrostatic barrier.
Then we analyze the size distribution at some selected locations in more detail. The neglect of dust drift in our modeling
does not allow us to rely on the macroscopic tail of the grain size distribution ($\sim 1$\,mm) on a par with the current
state of the art dust evolution models. But for the goals of the paper we need the very fact that grains are able to
overcome $\sim 10\,\mu$m size at some disk locations. This would mean the insignificance of the electrostatic barrier there.

The average grain mass at every disk location is defined as
\begin{equation}
\overline{m}=\left. \int mf(m)dm \middle/ \int f(m)dm \right.,
\end{equation}
where $f(m)$ is the grain mass distribution. In Figure~\ref{fig:1} we show the radial and vertical distribution of an equivalent  average grain radius (i.e., of
an equivalent radius of a grain with the average mass $\overline{m}$) after 0.9\,Myr since the start of dust coagulation for six cases:
\begin{figure*}[ht!]
\plotone{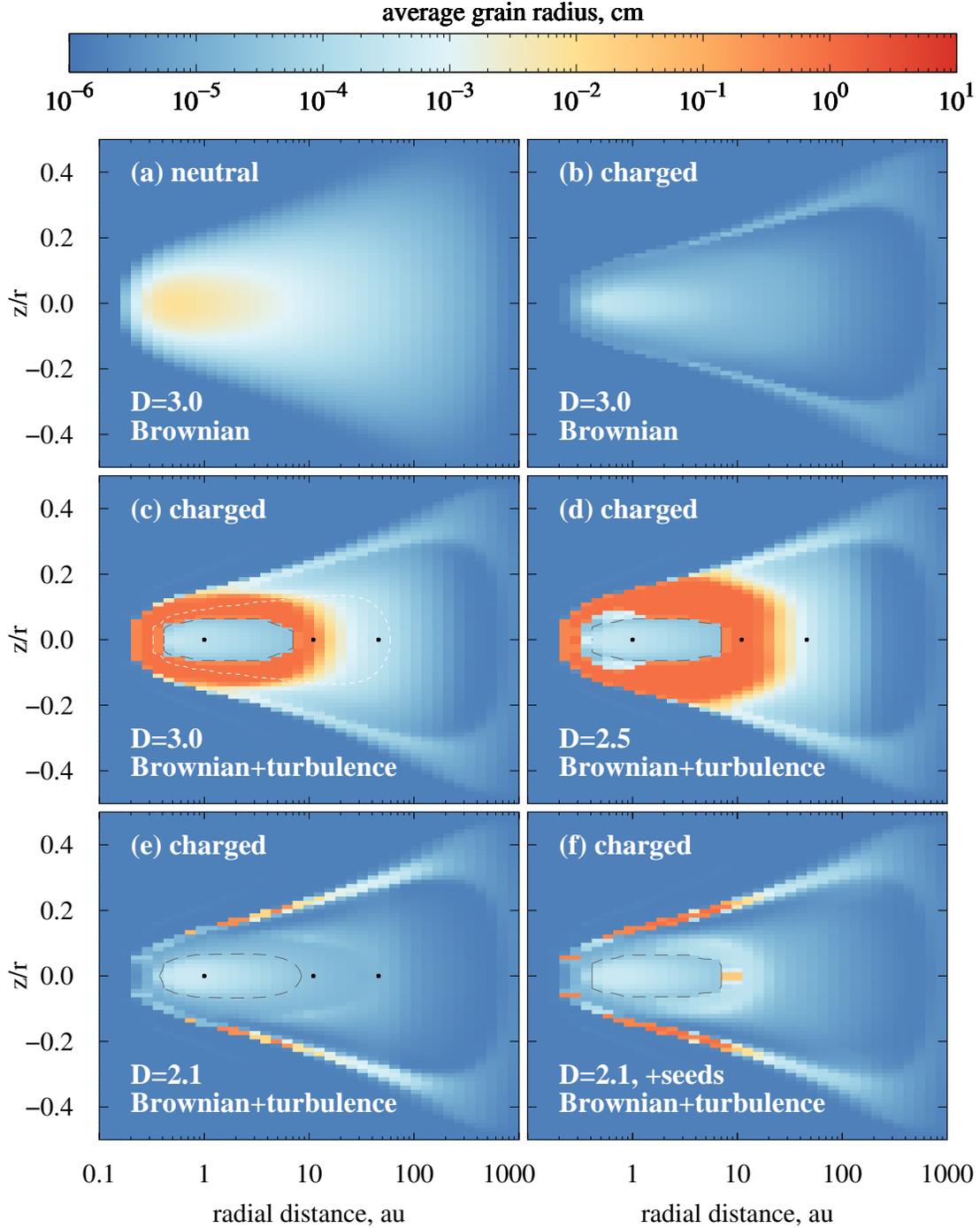} \caption{Equivalent average grain radius in different disk locations after 0.9~Myr of {\it in situ} coagulation.
Panels (a) and (b) show purely Brownian coagulation of neutral (a) and charged (b) grains. For other panels the
turbulence-induced velocities are also included. Panels (c)--(e) present the results for different
values of the fractal dimension of dust aggregates, $D=3.0, 2.5,$ and $2.1$. Panel (f) represents the case
from panel (e), but with initial size distribution contaminated by the artificially large $\approx 50\,\mu$m seed particles.
The long dashed lines show the location of a dead zone for $t=0.9\,$Myr ($\alpha_{\rm active}=10^{-3}, \alpha_{\rm
dead}=10^{-6}$), the short dashed line in panel (c) shows the location of the dead zone at $t=0$ (which is the
same for panels (d)--(f), but not shown on them). The black bullets indicate the locations for which the
data in Figure~\ref{fig:2} are presented. \label{fig:1}}
\end{figure*}
\begin{itemize}
 \item[(a)] purely Brownian coagulation of neutral dust;
 \item[(b)] purely Brownian coagulation of compact charged dust;
 \item[(c,d,e)] Brownian plus turbulence-induced coagulation of dust aggregates with fractal dimensions $D=3.0, 2.5$ and
     $2.1$;
 \item[(f)] case (e) for $D=2.1$, but with artificial presence of large $\approx 50\,\mu$m grains to simulate dust
     drift.
\end{itemize}

As noted above, the ratio of electrostatic-to-kinetic energy in Equation~(\ref{CF}) generally increases
with the grain size for Brownian motion, so the purely thermal coagulation of charged grains is inevitably stopped by the
electrostatic barrier at some limit size. We demonstrate this in panels~(a) and (b) of Figure~\ref{fig:1}. It is seen from
the panel~(b) that the Coulomb repulsion stops thermal dust coagulation at $\sim 1\,\mu$m throughout the whole disk volume.
Such small size limit is crucial for the subsequent dust growth as other sources of grain relative velocities are not strong
enough at this stage. The grain size on panel (a) of Figure~\ref{fig:1} may be overestimated as artificial growth is
possible if the number of size bins is not large enough (we divide the size range from $5\times10^{-7}$\,cm to $1$\,cm into
128 bins). However, for dust trapped behind the electrostatic barrier, this numerical effect is less important as there is a
physical mechanism blocking the growth of particles with sizes close to the initial ones. The fingerprint of the zero charge
surface, where the photoelectric and plasma charging are balanced, is seen in panel~(b) as a faint rim in the disk upper
layers. In this region, dust growth is almost not affected by the charging effects.

The consideration of turbulence-induced velocities solves the problem of electrostatic barrier for compact grains ($D=3.0$;
panel~(c)) everywhere in the disk except for the dead zone. The dead zone boundary at the moment $t=0.9$\,Myr is shown with
the long dashed line, while the short dashed line corresponds to the initial location of the dead zone. The dead zone shrinks with the
start of dust coagulation as the electron depletion onto dust grains become less important. This leads to the increase in
the ionization degree, which can exceed the critical value of $10^{-13}$ for the MRI development. Such behavior is strongly
affected by dust fragmentation as well as active and dead zone physics, which is surely quite simplistic in our model. This
stresses the need for separate study of self-consistent treatment of dust evolution and MRI development.

The macroscopic grain sizes denoted by the reddish color in Figure~\ref{fig:1} are not necessary attainable at a given
location due to several effects not considered in the presented modeling. They include the radial and vertical dust drift,
as well as fragmentation and compaction of the aggregates. Instead, the reddish color traces the disk regions where
the electrostatic barrier can be overcome and dust evolution is a charge-independent phenomenon.

The grain fractality ($D<3$) may play a double role in charged dust coagulation. First, it increases the collisional
cross section for a dust grain of given mass, thus boosting the coagulation rates. Second, it suppresses the
turbulence-induced dust velocities as they scale as $a^{(D-2)/2}$. Low values of $D$ may move forward the threshold grain
size where turbulence starts to dominate over the Brownian velocities. To explore the dependence of the dust coagulation on
dust fractality we considered two additional choices of $D=2.5$ and $ 2.1$.

One can see that the moderate fractality ($D=2.5$; Figure~\ref{fig:1}d) allows easier, charge-independent grain growth
outside the dead zone. This can not be said about the dead zone where almost all dust population is still locked behind the
electrostatic barrier. More fluffier aggregates ($D=2.1$, Figure~\ref{fig:1}e) are electrostatically blocked from
coagulation not only in the dead zone but almost everywhere in the disk, which is consistent with previous
studies~\citep{2009ApJ...698.1122O}. Such a severe barrier arose due to the fact that the electrostatic-to-kinetic energy ratio is
larger than unity at grain sizes smaller than those affected by the turbulence.

To check the dependence of our results on the improperly considered effects (radial and vertical drift of grown dust),
we artificially added grown aggregates with sizes $\approx 50\,\mu$m to the initial grain size distribution. Their mass
fraction is $\sim 10\%$ of the initial dust mass. This may simulate the charge-independent grain growth outside the dead
zone and subsequent inward drift. The results for $D=2.1$ are shown in Figure~\ref{fig:1}f. One can see that this helps to
overcome the electrostatic barrier just outside the outer dead zone boundary, but again not in the dead zone itself.

\begin{figure*}[ht!]
\plotone{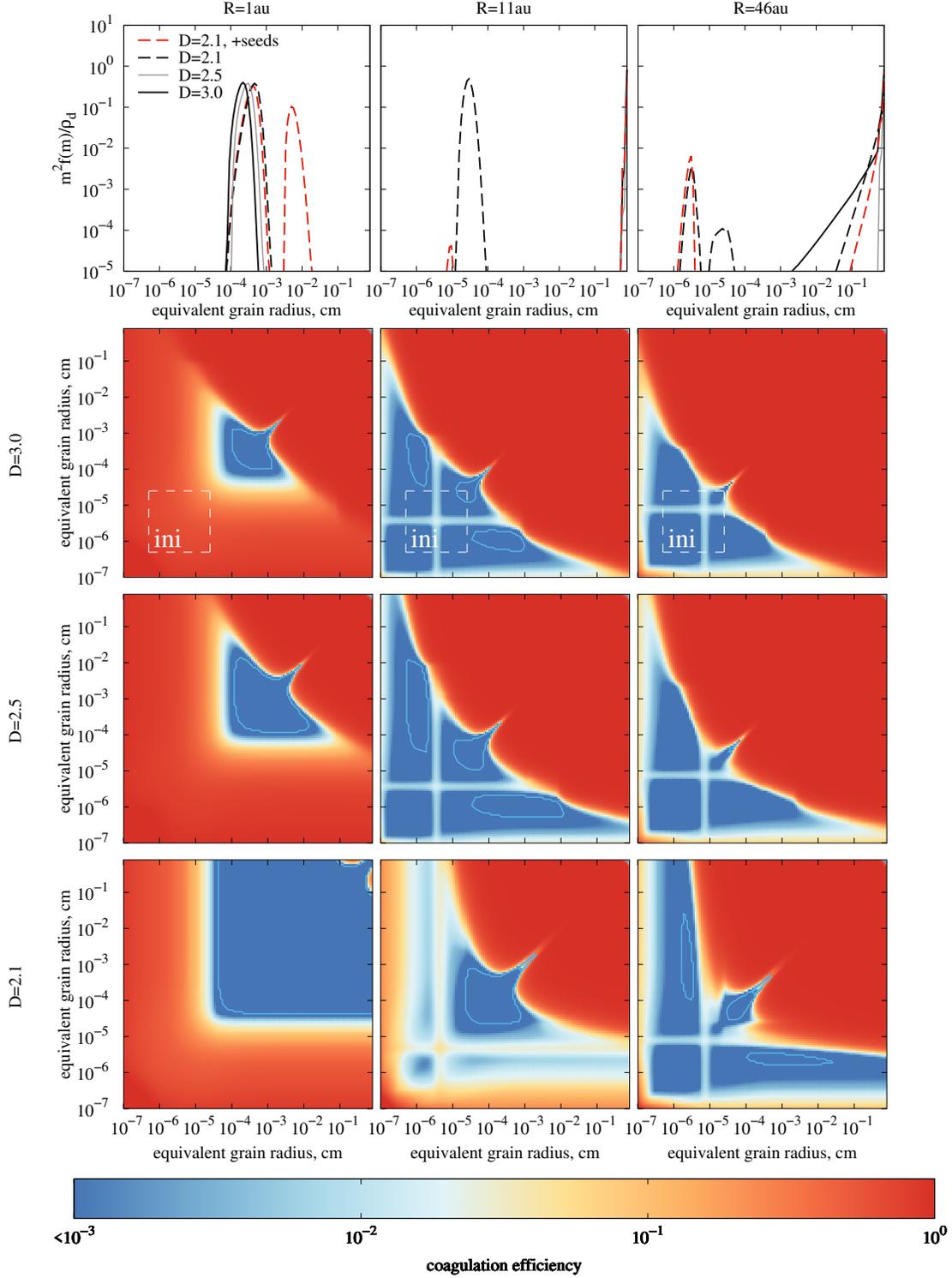} \caption{Grain size distribution at $t=0.9\,$Myr (upper row) and the charge-averaged Coulomb factor
$\overline{\mathcal{C}}(a_1,a_2)$ (coagulation efficiency) as a function of equivalent radius of colliding grains for fractal dimension
$D=3.0$ (second row), $D=2.5$ (third row), $D=2.1$ (bottom row). The columns correspond to different radial positions:
dead zone (1\,au; left column); active region near outside boundary of the dead zone (11\,au; middle column); and disk
periphery (46\,au; right column). The dashed squares marked with ``ini'' indicate the size range of the initial distribution (grains in the
initial distribution are assumed to be compact for all $D$). Blue contour lines confine the size domains with the 
coagulation efficiency of $\leq10^{-10}$. \label{fig:2}}
\end{figure*}

To have a closer look at the possibilities to overcome the electrostatic barrier we plotted in Figure~\ref{fig:2} the value
of the Coulomb factor integrated over the grain charge states (coagulation efficiency),
\begin{equation}\label{eq:coag_efficiency}
\overline{\mathcal{C}}(a_1,a_2)=\iint \mathcal{C}_{12}G_1 G_2 \,dQ_1dQ_2,
\end{equation}
where $G_i\equiv G(a_i,Q_i)$ is the Gaussian charge distribution of grains having the radius $a_i$. The
corresponding values of average grain charge and its dispersion are calculated similar to \citet{2015ARep...59..747A}. We
show  the results for three choices of fractal dimension $D=3.0,2.5,$ and 2.1 as well as three radial locations at $1,11,$ and $46$\,au
lying well inside the dead zone, just outside it and at the disk periphery, respectively.

Figure~\ref{fig:2} shows that the electrostatic barrier plays almost no role for coagulation of grains
with $a\gtrsim100~\mu$m (except for very fractal grains with $D=2.1$, located in the dead zone). This means that if large
``seed particles'' can be transported into the dead zones for cases $D=2.5$ and $3.0$, this should trigger the dust growth.
A similar effect should occur outside the dead zone at 11\,au for $D=2.1$ (see middle panel in the last row of
Figure~\ref{fig:2}). At that location, the coagulation within the ensemble of small dust is inhibited, but the coagulation
with larger particles is barrier-free. Seed particles added to that location absorb the small grains, which are otherwise
electrostatically ``locked'' (compare black and red dashed lines in the middle panel of the first row of
Figure~\ref{fig:2}).

\begin{figure}[ht!]
 \plotone{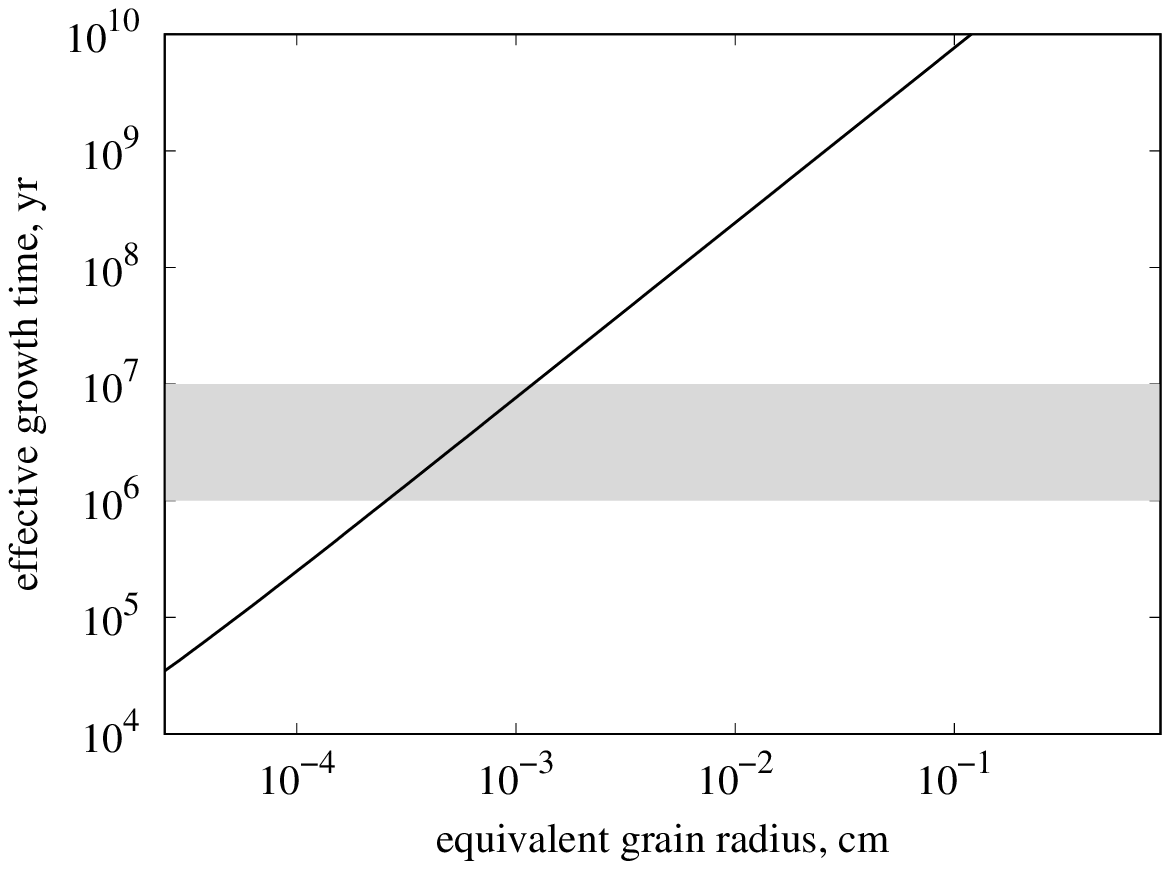} \caption{Effective timescale of dust coagulation due to turbulence-induced crossing of the
 zero-charge layer, plotted versus the equivalent grain radius. The horizontal gray stripe shows the range of
 typical lifetimes of protoplanetary disks. \label{fig:3}}
\end{figure}

In the vicinity of the zero-charge layer, dust grains can freely coagulate. The sustained turbulence ensures repetitive
passage of aerodynamically small grains through this zone, which provides a potential mechanism to overcome the
electrostatic barrier even for low fractality of $D=2.1$ (where the addition of $\sim50\mu m$ seed particles is not
efficient). To evaluate the significance of this mechanism, we notice that the size growth rate \citep[see, e.g.,][Section 3.3]{2016SSRv..205...41B} can be straightforwardly
generalized for fractal grains,
\begin{equation}
 \dot{a}=u_{12}\frac{\rho_{\rm d}}{\rho_{\rm s}}\frac{3}{D} \left(\frac{a}{a_0}\right)^{3-D},
\end{equation}
where $\rho_{\rm d}$ is the volume density of dust in the disk. The resulting charge-free coagulation timescale, $t_{\rm
coag}^0=a/\dot{a}$, should be modified to account for limited time grains staying in the zero-charge layer. With the vertical
thickness $\Delta z_0$ of the layer at the vertical position $z_0$, the ratio $\Delta z_0/z_0$ is an order-of-magnitude
estimate of the time fraction the grains are spending in the layer. We define $\Delta z_0 (a)$ as the height difference
where the average charge number of a grain changes from $+1$ to $-1$. The layer thickness shrinks for larger grains as the
vertical charge gradient increases with their size. In Figure~\ref{fig:3}, we plot the effective growth timescale,
$(z_0/\Delta z_0)t_{\rm coag}^0$, as a function of the equivalent grain radius. We see that for $a>10\mu m$ the timescale
becomes larger than the typical protoplanetary disk lifetime of $\sim 10^7$\,yr. Thus, even for $D=2.1$ the turbulent vertical stirring of dust appears to be inefficient in overcoming the electrostatic
barrier. 

To understand the shape of the ``inhibited zones'' (blue regions) in Figure~\ref{fig:2}, let us evaluate the ratio of the
contact energy of the electrostatic repulsion between two grains to the kinetic energy of their relative motion---according
to Equation~(\ref{CF}), this value determines the deviation of $\overline{\mathcal{C}}(a_1,a_2)$ from unity.
Given a Gaussian charge distribution, we expect the average coagulation rate to be drastically reduced when this ratio
(calculated for the average charge) is of the order of unity or larger. Considering a pair of grains with radii $a_1>a_2$,
this condition can be approximately presented in the following form:
\begin{equation}\label{condition}
\frac{\tilde\varphi_0|Z_1Z_2|/\tilde a_1}{1+c\alpha{\rm St}_0(m_0/m_{\rm g})\tilde a_1^{D-2}\tilde a_2^D}\gtrsim1.
\end{equation}
Here, $\tilde a_{1,2}=a_{1,2}/a_0$ are the grain radii normalized by the radius of a monomer, ${\rm St}_0=\Omega_{\rm
K}t_{\rm s0}~(\ll1)$ is the monomer Stokes number (expressed via the local Keplerian frequency $\Omega_{\rm K}\propto
R^{-3/2}$ and the local stopping time of the monomer $t_{\rm s0}\propto a_0\rho_{\rm g}^{-1}T^{-1/2}$), and $c$ is a
constant of the order of unity. The relative magnitude of the repulsion barrier is given by the product of charge numbers,
$Z_{1,2}=Q_{1,2}/e$, multiplied with the unit-charge electrostatic energy at the monomer surface (normalized to the thermal
energy), $\tilde\varphi_0=e^2/(a_0k_{\rm B}T)$.

As long as grains are small enough, $\tilde \varphi_0/\tilde a_i\gg1$, they are (typically) singly charged ($Z_i=-1$) due to
the induced-dipole attraction of plasma charges. Furthermore, for small grains the term in the denominator of
Equation~(\ref{condition}) is negligible. Therefore, the left-hand side is simply $\tilde \varphi_0/\tilde a_1\gg1$ and hence the
coagulation of such grains is completely inhibited. Of course, reducing the grain size further makes the grains neutral, but
this transition typically occurs when the size is about a nanometer, as one can see in the middle and right columns of
Figure~\ref{fig:2}. On the other hand, the left column represents the plasma regime with a strongly depleted electron
density (see Section~\ref{DP_regimes}), and therefore grains of the initial size distribution remain neutral in this case.

The transition to multiply charged grains at a given location occurs where $\tilde \varphi_0/\tilde a_i\simeq1$ (assuming
the electron depletion is not strong), which is revealed in Figure~\ref{fig:2} by the crossing lines in the inhibited zone.
The origin of these lines is obvious -- Equation~(\ref{condition}) in this case is marginally satisfied, and therefore the
coagulation rate is only moderately reduced. The further increase of the grain sizes leads to the average charges such that
the product $\tilde \varphi_0|Z_i|/\tilde a_i$ is a constant of the order of a few (its value is determined by the local
plasma regime). Hence, the numerator of Equation~(\ref{condition}) becomes proportional to $a_2T$. For sufficiently large
grains the second term in the denominator is dominant, and then the lhs starts rapidly decreasing. Taking into account the
scaling dependence of ${\rm St}_0$ on the radial position, we conclude that the electrostatic barrier is unimportant if the
sizes exceed a threshold determined from the relation $\alpha R^{3/2}\tilde a_1^{D-2}\tilde a_2^{D-1}=$~const. Except for
the proximity of the diagonal line $a_1=a_2$, where more detailed analysis is required, this relation well describes the
``outer'' boundaries of the inhibited zones in Figure~\ref{fig:2}.

\subsection{Dusty plasma regimes}\label{DP_regimes}
\begin{figure*}[ht!]
\plottwo{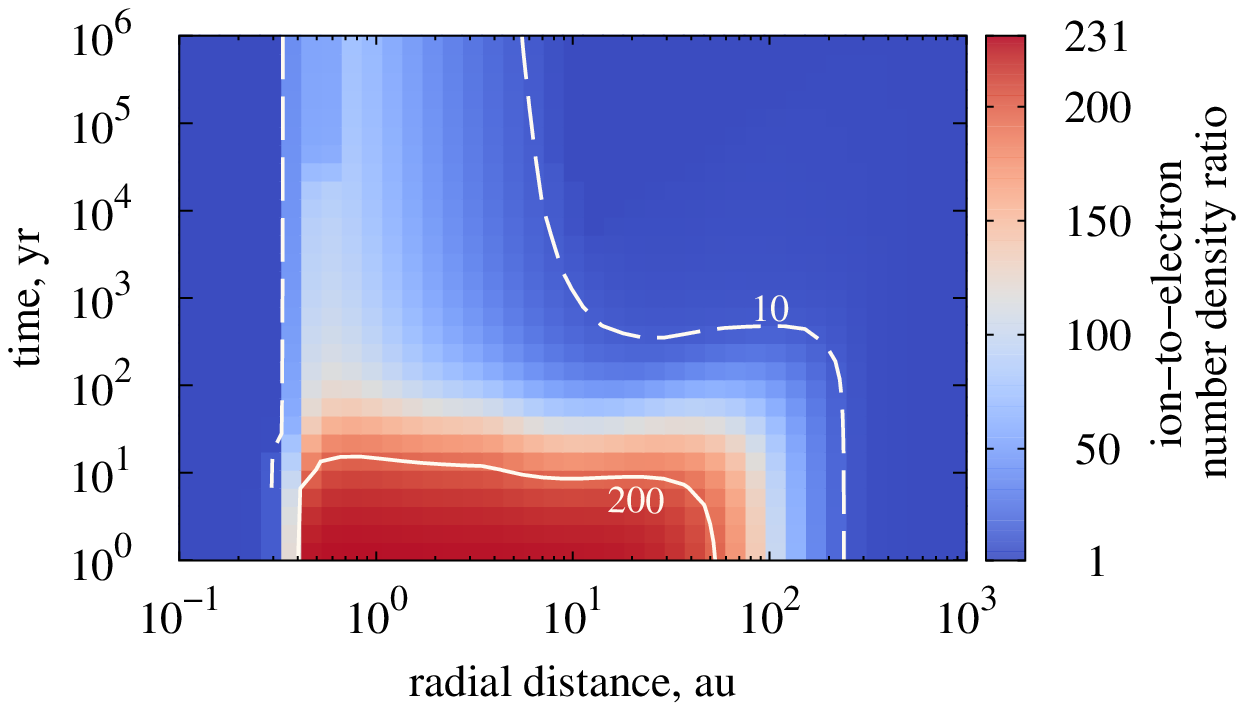}{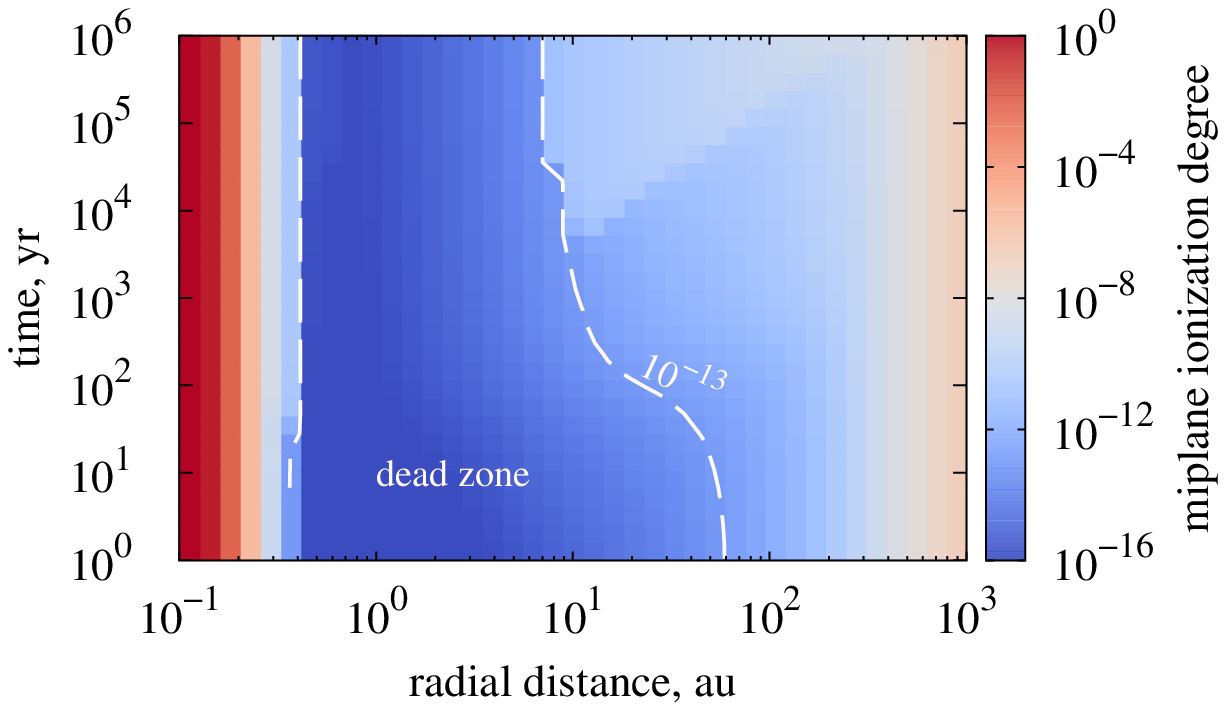} \caption{Left panel: the evolution of the ion-to-electron ratio $n_{\rm
i}/n_{\rm e}$ in the disk midplane for the case of non-fractal dust ($D=3.0$; corresponds to panel (c) in
Figure~\ref{fig:1}). Right panel: the evolution of the ionization degree $x_{\rm e}=n_{\rm e}/n_{\rm gas}$ in
the disk midplane for the same case. The dead zone with suppressed turbulence ($\alpha_{\rm dead}=10^{-6}$) is
defined as a region with $x_{\rm e}<10^{-13}$. Its size shrinks with dust evolution.  \label{fig:4}}
\end{figure*}
Generally, the coagulation may be stopped by the electrostatic barrier at either dust--ion or electron--ion state depending on
the underlying grain-grain relative velocities. The wealth of free electrons in the electron--ion plasma allows the most
negative grain charges (for given plasma temperature and ion mass) and, consequently, the hardest conditions for the grain
growth. In high-density low-ionization regions, where the ionization degree is comparable with the dust abundance, the
plasma state shifts to the dust--ion regime and grains become less charged as fewer electrons hit the dust. In the limit 
of dust--dust plasma, where both electrons and ions are severely depleted from gas, the average grain charge is near zero and
the electrostatic barrier disappears~\citep{2016ApJ...833...92I}. The largest disproportion in abundances of ions and
electrons,
\begin{equation}
 \left(\frac{n_{\rm i}}{n_{\rm e}}\right)_{\rm max}=\frac{s_{\rm e}}{s_{\rm i}}\sqrt{\frac{m_{\rm i}}{m_{\rm e}}}
\end{equation}
is achieved in dust--dust plasma. In the case of perfect electron and ion sticking ($s_{\rm e}=s_{\rm i}=1$) and N$_2$H$^+$
or HCO$^+$ being the dominant ion ($m_{\rm i}=29m_{\rm H}$), the maximum ion-to-electron number density ratio is
$\left(n_{\rm i}/n_{\rm e}\right)_{\rm max}=231$. The grain growth in absence of fragmentation changes the plasma state in
the direction from dust--dust state to dust--ion and then to electron--ion state. This also leads to the decrease of $n_{\rm
i}/n_{\rm e}$ down to unity and to the increase in the ionization degree.

In the left panel of Figure~\ref{fig:4}, we show the evolution of the ion-to-electron ratio $n_{\rm i}/n_{\rm e}$ in the
disk midplane for the case of non-fractal dust (corresponding to the case (c) in Figure~\ref{fig:1}). At the very beginning
of simulations the plasma is in the dust--dust state within 0.4--50\,au and $n_{\rm i}/n_{\rm e}>200$. However, it quickly
switches to the dust--ion and electron--ion state. In the right panel of Figure~\ref{fig:4}, we show the corresponding
ionization degree. The dashed isoline for $x_{\rm e}=10^{-13}$ demonstrates the evolution of the dead zone due to the dust
coagulation. By the end of simulations at $0.9$\,Myr the dead zone shrinks from 0.3--60\,au to 0.4--7\,au and coincides with
the electrostatic barrier region. While the electrostatic barrier may occur in both dust--ion and electron--ion plasma, for
non-fractal dust ($D=3.0$) and chosen disk parameters it happened in dust--ion regime (see also~\citet{2012A&A...538A.124I}).

\section{Discussion} \label{sec:discussion}

While protoplanetary disks are likely to have favorable conditions for coagulation of micrometer-size
particles into macroscopic pebbles, we still lack the clear understanding of this pathway. Dense midplane regions with mild
sources of relative grain velocities are thought to foster the rapid dust coagulation up to the sizes at which grains start
to experience the notable drag relative to the gas. This allows the grains to concentrate in dust traps and trigger the
subsequent formation of planetesimals and, eventually, planets. At the same time, small micrometer-size dust is persistent
along the entire protoplanetary disk evolution. It provides the key source of disk opacity in ultraviolet and optical range
and shields molecules in disk interiors from dissociation. The presence of small dust can be inferred from near-IR spectral
energy distributions~\citep{2001AJ....121.1512H, 2007ApJ...667..308C} and imaging~\citep{2018ApJ...863...44A}.

Sustaining large amounts of small dust can be explained by the replenishment of its population from larger grains due to the
fragmentation~\citep{2005A&A...434..971D}, or/and by extremely inefficient coagulation in micrometer-size
range due to the Coulomb repulsion. The important difference between these two alternatives 
is the role of turbulence. If the fragmentation is the dominant
mechanism of small dust replenishment, one may expect larger amounts of micron-size dust in disks with stronger turbulence.
On the opposite, strong turbulence makes the electrostatic barrier less effective in maintaining small dust population.
Thus, future measurements of non-thermal line widths may be crucial for our understanding of dust evolution in
protoplanetary disks~\citep{2017ApJ...843..150F, 2018ApJ...856..117F}.

\section{Conclusions}
Coagulation of small particles is a key process in protoplanetary disk evolution and formation of planets. This process is
controlled by a number of microphysical factors like sticking, bouncing, sintering and fragmentation as well as global dust
dynamics. In this paper we study the coagulation of grains charged due to photoelectric effect and the collection of free
electrons and ions. We solve the Smoluchowski equation coupled with grain charging and gas ionization equations to study the
conditions where the electrostatic barrier between like-charged grains can play an important role. As the
Coulomb repulsion is typically important for small $0.1-10\,\mu$m grains, which are not large enough to drift relative to the
gas, we neglect global dust dynamics. The simulations are done in the 2D vertical and radial extent of a typical
protoplanetary disk and account for the simple self-consistent co-evolution of the dead zone and dust inside it. We consider
three characteristic values of the fractal dimension of dust aggregates, $D=3.0, 2.5,$ and $2.1$, defining a
dependence of the grain mass on the size, $m\propto a^D$. Our conclusions can be summarized as follows:
\begin{enumerate}
  \item Small 0.1--10\,$\mu$m grains in protoplanetary disk interiors are sufficiently negatively charged to inhibit their mutual collisions.
  \item Sustained turbulence with $\alpha\gtrsim10^{-3}$ is necessary, but not sufficient, to overcome the electrostatic barrier between small grains, which makes the initial dust growth blocked in dead zones.
  \item Although the {\it mutual} coagulation of small particles in the dead zone is inhibited, their collisions with large grains ($>100\,\mu$m) having $D=2.5$ or $3.0$ are possible. Thus, large particles (drifting, e.g., from the outer disk) may serve as seeds for barrier-free coagulation in the dead zone.
  \item The coagulation of highly fractal dust with $D=2.1$ is blocked not only in the dead zone, but almost in the entire disk.
\end{enumerate}
Thus, the mutual electrostatic repulsion of small, $\mu$m-size grains can efficiently prevent their coagulation
in regions where large dust is absent. This can serve as an alternative (to fragmentation) mechanism explaining the presence
of small dust in disk atmosphere and outer regions.

\acknowledgments

We thank Kedron Silsbee and an anonymous referee for constructive critical comments, improving the manuscript. 
The research was supported by the Russian Science Foundation (project No. 17-12-01441; Section 3) and by the
Russian Foundation for Basic Research grant 18-52-52006.

\bibliographystyle{aasjournal}
\bibliography{refs}

\begin{thebibliography}{}
\expandafter\ifx\csname natexlab\endcsname\relax\def\natexlab#1{#1}\fi
\providecommand{\url}[1]{\href{#1}{#1}}

\bibitem[{{Akimkin}(2017)}]{2017ASPC..510...63A}
{Akimkin}, V. 2017, in Astronomical Society of the Pacific Conference Series,
  Vol. 510, Stars: From Collapse to Collapse, ed. Y.~Y. {Balega}, D.~O.
  {Kudryavtsev}, I.~I. {Romanyuk}, \& I.~A. {Yakunin}, 63

\bibitem[{{Akimkin}(2015)}]{2015ARep...59..747A}
{Akimkin}, V.~V. 2015, Astronomy Reports, 59, 747

\bibitem[{{Akimkin} {et~al.}(2017){Akimkin}, {Kirsanova}, {Pavlyuchenkov}, \&
  {Wiebe}}]{2017MNRAS.469..630A}
{Akimkin}, V.~V., {Kirsanova}, M.~S., {Pavlyuchenkov}, Y.~N., \& {Wiebe}, D.~S.
  2017, \mnras, 469, 630

\bibitem[{Armitage(2018)}]{Armitage2018}
Armitage, P.~J. 2018, A Brief Overview of Planet Formation, ed. H.~J. Deeg \&
  J.~A. Belmonte (Cham: Springer International Publishing), 2185--2203

\bibitem[{{Avenhaus} {et~al.}(2018){Avenhaus}, {Quanz}, {Garufi}, {Perez},
  {Casassus}, {Pinte}, {Bertrang}, {Caceres}, {Benisty}, \&
  {Dominik}}]{2018ApJ...863...44A}
{Avenhaus}, H., {Quanz}, S.~P., {Garufi}, A., {et~al.} 2018, \apj, 863, 44

\bibitem[{{Bai} \& {Goodman}(2009)}]{2009ApJ...701..737B}
{Bai}, X.-N., \& {Goodman}, J. 2009, \apj, 701, 737

\bibitem[{{Bakes} \& {Tielens}(1994)}]{1994ApJ...427..822B}
{Bakes}, E.~L.~O., \& {Tielens}, A.~G.~G.~M. 1994, \apj, 427, 822

\bibitem[{{Barranco} \& {Goodman}(1998)}]{1998ApJ...504..207B}
{Barranco}, J.~A., \& {Goodman}, A.~A. 1998, \apj, 504, 207

\bibitem[{{Birnstiel} {et~al.}(2010){Birnstiel}, {Dullemond}, \&
  {Brauer}}]{2010A&A...513A..79B}
{Birnstiel}, T., {Dullemond}, C.~P., \& {Brauer}, F. 2010, \aap, 513, A79

\bibitem[{{Birnstiel} {et~al.}(2016){Birnstiel}, {Fang}, \&
  {Johansen}}]{2016SSRv..205...41B}
{Birnstiel}, T., {Fang}, M., \& {Johansen}, A. 2016, \ssr, 205, 41

\bibitem[{{Blum}(2010)}]{2010RAA....10.1199B}
{Blum}, J. 2010, Research in Astronomy and Astrophysics, 10, 1199

\bibitem[{{Boersma} {et~al.}(2018){Boersma}, {Bregman}, \&
  {Allamandola}}]{2018ApJ...858...67B}
{Boersma}, C., {Bregman}, J., \& {Allamandola}, L.~J. 2018, \apj, 858, 67

\bibitem[{{Bohren} \& {Hunt}(1977)}]{1977CaJPh..55.1930B}
{Bohren}, C.~F., \& {Hunt}, A.~J. 1977, Canadian Journal of Physics, 55, 1930

\bibitem[{{Caselli} {et~al.}(2002){Caselli}, {Benson}, {Myers}, \&
  {Tafalla}}]{2002ApJ...572..238C}
{Caselli}, P., {Benson}, P.~J., {Myers}, P.~C., \& {Tafalla}, M. 2002, \apj,
  572, 238

\bibitem[{{Chokshi} {et~al.}(1993){Chokshi}, {Tielens}, \&
  {Hollenbach}}]{1993ApJ...407..806C}
{Chokshi}, A., {Tielens}, A.~G.~G.~M., \& {Hollenbach}, D. 1993, \apj, 407, 806

\bibitem[{{Cieza} {et~al.}(2007){Cieza}, {Padgett}, {Stapelfeldt}, {Augereau},
  {Harvey}, {Evans}, {Mer{\'{\i}}n}, {Koerner}, {Sargent}, {van Dishoeck},
  {Allen}, {Blake}, {Brooke}, {Chapman}, {Huard}, {Lai}, {Mundy}, {Myers},
  {Spiesman}, \& {Wahhaj}}]{2007ApJ...667..308C}
{Cieza}, L., {Padgett}, D.~L., {Stapelfeldt}, K.~R., {et~al.} 2007, \apj, 667,
  308

\bibitem[{{Corlin}(1938)}]{1938ZA.....15..239C}
{Corlin}, A. 1938, \zap, 15, 239

\bibitem[{{DeFrees} {et~al.}(1993){DeFrees}, {Miller}, {Talbi}, {Pauzat}, \&
  {Ellinger}}]{1993ApJ...408..530D}
{DeFrees}, D.~J., {Miller}, M.~D., {Talbi}, D., {Pauzat}, F., \& {Ellinger}, Y.
  1993, \apj, 408, 530

\bibitem[{{Desch} \& {Cuzzi}(2000)}]{2000Icar..143...87D}
{Desch}, S.~J., \& {Cuzzi}, J.~N. 2000, \icarus, 143, 87

\bibitem[{{Draine}(1980)}]{1980ApJ...241.1021D}
{Draine}, B.~T. 1980, \apj, 241, 1021

\bibitem[{{Draine} \& {Salpeter}(1979)}]{1979ApJ...231...77D}
{Draine}, B.~T., \& {Salpeter}, E.~E. 1979, \apj, 231, 77

\bibitem[{{Draine} \& {Sutin}(1987)}]{1987ApJ...320..803D}
{Draine}, B.~T., \& {Sutin}, B. 1987, \apj, 320, 803

\bibitem[{{Dudorov} \& {Khaibrakhmanov}(2014)}]{2014Ap&SS.352..103D}
{Dudorov}, A.~E., \& {Khaibrakhmanov}, S.~A. 2014, \apss, 352, 103

\bibitem[{{Dullemond} \& {Dominik}(2005)}]{2005A&A...434..971D}
{Dullemond}, C.~P., \& {Dominik}, C. 2005, \aap, 434, 971

\bibitem[{{Dzyurkevich} {et~al.}(2013){Dzyurkevich}, {Turner}, {Henning}, \&
  {Kley}}]{2013ApJ...765..114D}
{Dzyurkevich}, N., {Turner}, N.~J., {Henning}, T., \& {Kley}, W. 2013, \apj,
  765, 114

\bibitem[{{Feuerbacher} {et~al.}(1973){Feuerbacher}, {Willis}, \&
  {Fitton}}]{1973ApJ...181..101F}
{Feuerbacher}, B., {Willis}, R.~F., \& {Fitton}, B. 1973, \apj, 181, 101

\bibitem[{{Flaherty} {et~al.}(2018){Flaherty}, {Hughes}, {Teague}, {Simon},
  {Andrews}, \& {Wilner}}]{2018ApJ...856..117F}
{Flaherty}, K.~M., {Hughes}, A.~M., {Teague}, R., {et~al.} 2018, \apj, 856, 117

\bibitem[{{Flaherty} {et~al.}(2017){Flaherty}, {Hughes}, {Rose}, {Simon}, {Qi},
  {Andrews}, {K{\'o}sp{\'a}l}, {Wilner}, {Chiang}, {Armitage}, \&
  {Bai}}]{2017ApJ...843..150F}
{Flaherty}, K.~M., {Hughes}, A.~M., {Rose}, S.~C., {et~al.} 2017, \apj, 843,
  150

\bibitem[{{Fortov} {et~al.}(2005){Fortov}, {Ivlev}, {Khrapak}, {Khrapak}, \&
  {Morfill}}]{2005PhR...421....1F}
{Fortov}, V.~E., {Ivlev}, A.~V., {Khrapak}, S.~A., {Khrapak}, A.~G., \&
  {Morfill}, G.~E. 2005, Physics Reports, 421, 1

\bibitem[{{Gail} \& {Sedlmayr}(1979)}]{1979A&A....77..165G}
{Gail}, H.~P., \& {Sedlmayr}, E. 1979, \aap, 77, 165

\bibitem[{{Goertz}(1989)}]{1989RvGeo..27..271G}
{Goertz}, C.~K. 1989, Reviews of Geophysics, 27, 271

\bibitem[{{Haisch} {et~al.}(2001{\natexlab{a}}){Haisch}, {Lada}, \&
  {Lada}}]{2001ApJ...553L.153H}
{Haisch}, Jr., K.~E., {Lada}, E.~A., \& {Lada}, C.~J. 2001{\natexlab{a}},
  \apjl, 553, L153

\bibitem[{{Haisch} {et~al.}(2001{\natexlab{b}}){Haisch}, {Lada}, {Pi{\~n}a},
  {Telesco}, \& {Lada}}]{2001AJ....121.1512H}
{Haisch}, Jr., K.~E., {Lada}, E.~A., {Pi{\~n}a}, R.~K., {Telesco}, C.~M., \&
  {Lada}, C.~J. 2001{\natexlab{b}}, \aj, 121, 1512

\bibitem[{{Harper} {et~al.}(2018){Harper}, {Helling}, \&
  {Dufek}}]{0004-637X-867-2-123}
{Harper}, J.~M., {Helling}, C., \& {Dufek}, J. 2018, \apj, 867, 123

\bibitem[{{Hopkins} \& {Squire}(2018)}]{2018MNRAS.479.4681H}
{Hopkins}, P.~F., \& {Squire}, J. 2018, \mnras, 479, 4681

\bibitem[{{Horanyi} \& {Goertz}(1990)}]{1990ApJ...361..155H}
{Horanyi}, M., \& {Goertz}, C.~K. 1990, \apj, 361, 155

\bibitem[{{Ilgner}(2012)}]{2012A&A...538A.124I}
{Ilgner}, M. 2012, \aap, 538, A124

\bibitem[{{Ilgner} \& {Nelson}(2006)}]{2006A&A...445..205I}
{Ilgner}, M., \& {Nelson}, R.~P. 2006, \aap, 445, 205

\bibitem[{{Ivlev} {et~al.}(2016){Ivlev}, {Akimkin}, \&
  {Caselli}}]{2016ApJ...833...92I}
{Ivlev}, A.~V., {Akimkin}, V.~V., \& {Caselli}, P. 2016, \apj, 833, 92

\bibitem[{{Ivlev} {et~al.}(2015){Ivlev}, {Padovani}, {Galli}, \&
  {Caselli}}]{Ivlev2015}
{Ivlev}, A.~V., {Padovani}, M., {Galli}, D., \& {Caselli}, P. 2015, \apj, 812,
  135

\bibitem[{{Jones} {et~al.}(1996){Jones}, {Tielens}, \&
  {Hollenbach}}]{1996ApJ...469..740J}
{Jones}, A.~P., {Tielens}, A.~G.~G.~M., \& {Hollenbach}, D.~J. 1996, \apj, 469,
  740

\bibitem[{{Jung}(1937)}]{1937AN....263..425J}
{Jung}, B. 1937, Astronomische Nachrichten, 263, 425

\bibitem[{{Katushkina} {et~al.}(2018){Katushkina}, {Alexashov}, {Gvaramadze},
  \& {Izmodenov}}]{2018MNRAS.473.1576K}
{Katushkina}, O.~A., {Alexashov}, D.~B., {Gvaramadze}, V.~V., \& {Izmodenov},
  V.~V. 2018, \mnras, 473, 1576

\bibitem[{{Keto} {et~al.}(2015){Keto}, {Caselli}, \&
  {Rawlings}}]{2015MNRAS.446.3731K}
{Keto}, E., {Caselli}, P., \& {Rawlings}, J. 2015, \mnras, 446, 3731

\bibitem[{{Kocifaj} \& {Kla{\v{c}}ka}(2012)}]{2012OptL...37..265K}
{Kocifaj}, M., \& {Kla{\v{c}}ka}, J. 2012, Optics Letters, 37, 265

\bibitem[{{Kocifaj} {et~al.}(2012){Kocifaj}, {Kla{\v{c}}ka}, {Videen}, \&
  {Koh{\'u}t}}]{2012JQSRT.113.2561K}
{Kocifaj}, M., {Kla{\v{c}}ka}, J., {Videen}, G., \& {Koh{\'u}t}, I. 2012,
  Journal of Quantitative Spectroscopy and Radiative Transfer, 113, 2561

\bibitem[{{Kopnin} {et~al.}(2011){Kopnin}, {Morozova}, \&
  {Popel}}]{2011AIPC.1397..249K}
{Kopnin}, S.~I., {Morozova}, T.~I., \& {Popel}, S.~I. 2011, in American
  Institute of Physics Conference Series, Vol. 1397, American Institute of
  Physics Conference Series, ed. V.~Y. {Nosenko}, P.~K. {Shukla}, M.~H.
  {Thoma}, \& H.~M. {Thomas}, 249--250

\bibitem[{Langhoff(1996)}]{doi:10.1021/jp952074g}
Langhoff, S.~R. 1996, The Journal of Physical Chemistry, 100, 2819

\bibitem[{{Lee} {et~al.}(2017){Lee}, {Hopkins}, \&
  {Squire}}]{2017MNRAS.469.3532L}
{Lee}, H., {Hopkins}, P.~F., \& {Squire}, J. 2017, \mnras, 469, 3532

\bibitem[{{Lefevre}(1975)}]{1975A&A....41..437L}
{Lefevre}, J. 1975, \aap, 41, 437

\bibitem[{{Marshall} {et~al.}(2005){Marshall}, {Sauke}, \&
  {Cuzzi}}]{2005GeoRL..3211202M}
{Marshall}, J.~R., {Sauke}, T.~B., \& {Cuzzi}, J.~N. 2005, \grl, 32, L11202

\bibitem[{{Mathews}(1967)}]{1967ApJ...147..965M}
{Mathews}, W.~G. 1967, \apj, 147, 965

\bibitem[{{Matthews} {et~al.}(2013){Matthews}, {Shotorban}, \&
  {Hyde}}]{2013ApJ...776..103M}
{Matthews}, L.~S., {Shotorban}, B., \& {Hyde}, T.~W. 2013, \apj, 776, 103

\bibitem[{{McKee} {et~al.}(1987){McKee}, {Hollenbach}, {Seab}, \&
  {Tielens}}]{1987ApJ...318..674M}
{McKee}, C.~F., {Hollenbach}, D.~J., {Seab}, G.~C., \& {Tielens}, A.~G.~G.~M.
  1987, \apj, 318, 674

\bibitem[{{Mendis} \& {Rosenberg}(1994)}]{1994ARA&A..32..419M}
{Mendis}, D.~A., \& {Rosenberg}, M. 1994, \araa, 32, 419

\bibitem[{{Nesterenok}(2018)}]{2018Ap&SS.363..151N}
{Nesterenok}, A.~V. 2018, \apss, 363, 151

\bibitem[{{Okuzumi}(2009)}]{2009ApJ...698.1122O}
{Okuzumi}, S. 2009, \apj, 698, 1122

\bibitem[{{Okuzumi} {et~al.}(2011{\natexlab{a}}){Okuzumi}, {Tanaka},
  {Takeuchi}, \& {Sakagami}}]{2011ApJ...731...95O}
{Okuzumi}, S., {Tanaka}, H., {Takeuchi}, T., \& {Sakagami}, M.-a.
  2011{\natexlab{a}}, \apj, 731, 95

\bibitem[{{Okuzumi} {et~al.}(2011{\natexlab{b}}){Okuzumi}, {Tanaka},
  {Takeuchi}, \& {Sakagami}}]{2011ApJ...731...96O}
---. 2011{\natexlab{b}}, \apj, 731, 96

\bibitem[{{Ormel} \& {Cuzzi}(2007)}]{2007A&A...466..413O}
{Ormel}, C.~W., \& {Cuzzi}, J.~N. 2007, \aap, 466, 413

\bibitem[{{Padovani} {et~al.}(2018){Padovani}, {Ivlev}, {Galli}, \&
  {Caselli}}]{2018A&A...614A.111P}
{Padovani}, M., {Ivlev}, A.~V., {Galli}, D., \& {Caselli}, P. 2018, \aap, 614,
  A111

\bibitem[{{Pedersen} \& {G{\'o}mez de Castro}(2011)}]{2011ApJ...740...77P}
{Pedersen}, A., \& {G{\'o}mez de Castro}, A.~I. 2011, \apj, 740, 77

\bibitem[{{Pfalzner} {et~al.}(2014){Pfalzner}, {Steinhausen}, \&
  {Menten}}]{2014ApJ...793L..34P}
{Pfalzner}, S., {Steinhausen}, M., \& {Menten}, K. 2014, \apjl, 793, L34

\bibitem[{{Pineda} {et~al.}(2011){Pineda}, {Goodman}, {Arce}, {Caselli},
  {Longmore}, \& {Corder}}]{2011ApJ...739L...2P}
{Pineda}, J.~E., {Goodman}, A.~A., {Arce}, H.~G., {et~al.} 2011, \apjl, 739, L2

\bibitem[{{Poppe} {et~al.}(2000){Poppe}, {Blum}, \&
  {Henning}}]{2000ApJ...533..472P}
{Poppe}, T., {Blum}, J., \& {Henning}, T. 2000, \apj, 533, 472

\bibitem[{{Sano} {et~al.}(2000){Sano}, {Miyama}, {Umebayashi}, \&
  {Nakano}}]{2000ApJ...543..486S}
{Sano}, T., {Miyama}, S.~M., {Umebayashi}, T., \& {Nakano}, T. 2000, \apj, 543,
  486

\bibitem[{{Scalo}(1977)}]{1977A&A....55..253S}
{Scalo}, J.~M. 1977, \aap, 55, 253

\bibitem[{{Semenov} {et~al.}(2004){Semenov}, {Wiebe}, \&
  {Henning}}]{2004A&A...417...93S}
{Semenov}, D., {Wiebe}, D., \& {Henning}, T. 2004, \aap, 417, 93

\bibitem[{{Shchekinov}(2007)}]{2007A&AT...26..227S}
{Shchekinov}, Y. 2007, Astronomical and Astrophysical Transactions, 26, 227

\bibitem[{{Sicilia-Aguilar} {et~al.}(2006){Sicilia-Aguilar}, {Hartmann},
  {Calvet}, {Megeath}, {Muzerolle}, {Allen}, {D'Alessio}, {Mer{\'{\i}}n},
  {Stauffer}, {Young}, \& {Lada}}]{2006ApJ...638..897S}
{Sicilia-Aguilar}, A., {Hartmann}, L., {Calvet}, N., {et~al.} 2006, \apj, 638,
  897

\bibitem[{{Simpson}(1978)}]{1978Ap&SS..57..381S}
{Simpson}, I.~C. 1978, \apss, 57, 381

\bibitem[{{Simpson} {et~al.}(1979){Simpson}, {Simons}, \&
  {Williams}}]{1979Ap&SS..61...65S}
{Simpson}, J.~C., {Simons}, S., \& {Williams}, I.~P. 1979, \apss, 61, 65

\bibitem[{{Singh} \& {Mazza}(2018)}]{2018PhRvE..97b2904S}
{Singh}, C., \& {Mazza}, M.~G. 2018, \pre, 97, 022904

\bibitem[{{Spitzer}(1941)}]{1941ApJ....93..369S}
{Spitzer}, Jr., L. 1941, \apj, 93, 369

\bibitem[{{Thi} {et~al.}(2019){Thi}, {Lesur}, {Woitke}, {Kamp}, {Rab}, \&
  {Carmona}}]{2019A&A...632A..44T}
{Thi}, W.~F., {Lesur}, G., {Woitke}, P., {et~al.} 2019, \aap, 632, A44

\bibitem[{{Tielens} {et~al.}(1994){Tielens}, {McKee}, {Seab}, \&
  {Hollenbach}}]{1994ApJ...431..321T}
{Tielens}, A.~G.~G.~M., {McKee}, C.~F., {Seab}, C.~G., \& {Hollenbach}, D.~J.
  1994, \apj, 431, 321

\bibitem[{{Umebayashi} \& {Nakano}(1980)}]{1980PASJ...32..405U}
{Umebayashi}, T., \& {Nakano}, T. 1980, \pasj, 32, 405

\bibitem[{{Umebayashi} \& {Nakano}(2009)}]{2009ApJ...690...69U}
---. 2009, \apj, 690, 69

\bibitem[{{Watson}(1972)}]{1972ApJ...176..103W}
{Watson}, W.~D. 1972, \apj, 176, 103

\bibitem[{{Waxman} \& {Draine}(2000)}]{2000ApJ...537..796W}
{Waxman}, E., \& {Draine}, B.~T. 2000, \apj, 537, 796

\bibitem[{{Weingartner}(2004)}]{2004ASPC..309..453W}
{Weingartner}, J.~C. 2004, in Astrophysics of Dust, Vol. 309, 453

\bibitem[{{Weingartner} \& {Draine}(1999)}]{1999ApJ...517..292W}
{Weingartner}, J.~C., \& {Draine}, B.~T. 1999, \apj, 517, 292

\bibitem[{{Weingartner} \& {Draine}(2001)}]{2001ApJS..134..263W}
---. 2001, \apjs, 134, 263

\bibitem[{{Williams} \& {Cieza}(2011)}]{2011ARA&A..49...67W}
{Williams}, J.~P., \& {Cieza}, L.~A. 2011, \araa, 49, 67

\bibitem[{{Zhukovska} {et~al.}(2018){Zhukovska}, {Henning}, \&
  {Dobbs}}]{2018ApJ...857...94Z}
{Zhukovska}, S., {Henning}, T., \& {Dobbs}, C. 2018, \apj, 857, 94

\end{thebibliography}

\end{document}